\documentclass[aps,prb,twocolumn,superscriptaddress,showpacs,showkeys,floatfix]{revtex4}

\usepackage{graphicx,color}
\graphicspath{{figs/}}
\begin{document}
\title{Why twisting angles are diverse in graphene Moir{\'e} patterns?}
\author{Jin-Wu~Jiang}
    \affiliation{Institute of Structural Mechanics, Bauhaus-University Weimar, Marienstr. 15, D-99423 Weimar, Germany}
\author{Bing-Shen~Wang}
    \affiliation{State Key Laboratory of Semiconductor Superlattice and Microstructure and Institute of Semiconductor, Chinese Academy of Sciences, Beijing 100083, China}
\author{Timon~Rabczuk}
    \affiliation{Institute of Structural Mechanics, Bauhaus-University Weimar, Marienstr. 15, D-99423 Weimar, Germany}

\date{\today}
\begin{abstract}
The interlayer energy of the twisting bilayer graphene is investigated by the molecular mechanics method using both the registry-dependent potential and the Lennard-Jones potential. Both potentials show that the interlayer energy is independent of the twisting angle $\theta$, except in the two boundary regions $\theta\approx 0$ or $60^{\circ}$, where the interlayer energy is proportional to the square of the twisting arc length. The calculation results are successfully interpreted by a single atom model. An important information from our findings is that, from the energy point of view, there is no preference for the twisting angle in the experimental bilayer graphene samples, which actually explains the diverse twisting angles in the experiment.
\end{abstract}

\pacs{61.48.Gh, 61.72.Nn}
\keywords{Moir{\'e} patterns, twisting stacking fault, bilayer graphene, interlayer potential}
\maketitle

\section{Introduction}
The electronic band structure in single-layer graphene is well-known for its linear Dirac cones around K points in the Brillouin zone. This Dirac electron can be well described by a single $\pi$-orbital tight-binding model.\cite{PartoensB} Different from the single-layer graphene, a parabolic electronic dispersion was predicted theoretically for the Bernal-stacking bilayer graphene (BLG).\cite{PartoensB,McCannE} However, several experimental groups have observed the Dirac-like electron in BLG through direct or indirect measurements, which is attributed to the twisting defect in the BLG.\cite{ZhouSY,BergerC,SadowskiML,SprinkleM2009,LuicanA,HicksJ,Santos,Santos2012,LatilS,HassJ,BrownL} The twisting defect was also found to be responsible for the Van Hove singularities,\cite{LiG} the flat bands,\cite{MorellES} and the charge redistribution\cite{MorellES2011prb} in the BLG. The twisting pattern can be observed by the scanning tunneling microscopy\cite{VarchonF,HassJ,FloresM}, high-resolution transmission electron microscopy\cite{WarnerJH,JasinskiJB}, and atomic force microscopy.\cite{CarozoV} Besides electronic properties, it has been widely shown that single-layer graphene and BLG have peculiar mechanical and lattice properties.\cite{LeeCG,JiangJW2008,JiangJW2009,JiangJW2010,ShiX,TanPH} There is increasing interest in studying possible effects from the twisting pattern on the lattice properties of the BLG.\cite{PoncharalP,CarozoV,KimK,OhtaT,JiangJW2012} The most recent experiment demonstrates that the Raman spectrum strongly depends on the twisting angle of the Moir\'{e} pattern in BLG.\cite{KimK}

In above studies of various properties of the twisting BLG, a fundamental issue is to prepare the twisting angle of the BLG sample. The observed twisting angles are diverse in existing experiments. For example, in Ref.~\onlinecite{LuicanA}, Luican {\it et.al} obtained a twisting angle around $21.8^{\circ}$ or $3.5^{\circ}$. In Ref.~\onlinecite{BrownL}, Brown {\it et.al} have observed a broad distribution of the twisting angle around $29^{\circ}$, $24^{\circ}$, $17^{\circ}$, $12^{\circ}$, and $5^{\circ}$. The twisting angle is measured to be $4^{\circ}$ in Ref.~\onlinecite{WarnerJH}. Now, a fundamental question arises: Is there any preference for the twisting angle in the BLG sample? Should the experiment always observes a commensurable angle in the twisting BLG? The present work studies the interlayer interaction in the twisting BLG to shed some light on these questions from the energy point of view.

It has been a great challenge for theoretician to calculate the interlayer interaction in layered structures like BLG. To calculate the atomic energy, one may apply either density functional theory (DFT) or an empirical potential. In the empirical potential, the van der Waals (vdW) interaction in such layered structure is usually described by a Lennard-Jones potential. However, on the one hand, the interlayer interaction is long-range, so the standard DFT approach can not describe it, because the DFT is based on a local density approximation or a generalized gradient density approximation.\cite{WuX} On the other hand, it has been pointed out that the vdW potential itself is not sufficient to describe the interlayer energy, especially for the interlayer shearing movement, because the shearing is dominant by the $\pi$-overlap between different layers\cite{JeonGS,JiangJW2008prb} and the registry matching plays an important role.\cite{KolmogorovAN2000} The shearing property calculated from a pure vdW potential is one order smaller than the experimental value in such layered structure.\cite{KolmogorovAN2005,LebedevaIVpccp}

There are mainly two solutions for this issue. The first method is to develop a vdW-corrected DFT (DFT-D) approach, where the long-range interaction is described directly by the vdW potential\cite{GirifalcoLA,RydbergH,HasegawaM,OrtmannF,ErshovaOV} or is included through a density-density interaction in the DFT scheme\cite{DionM}. The second method is to include the $\pi$-overlap through some empirical potential terms with empirical parameters fitted to experiment or DFT-D results.\cite{KolmogorovAN2000,KolmogorovAN2004,KolmogorovAN2005,ZakharchenkoKV,HodO,MaromN,LebedevaIVjcp,LebedevaIVpccp} As pointed out by Girifalco and Hodak in 2002, these two methods are actually related to each other and should give the same results for interlayer interaction if they are applied properly.\cite{GirifalcoLA} In present study, the registry-dependent empirical interlayer potential will be applied in the calculation of the interlayer interaction, as long as we aim at simulations for large systems.

\begin{figure}[htpb]
  \begin{center}
    \scalebox{1.0}[1.0]{\includegraphics[width=8cm]{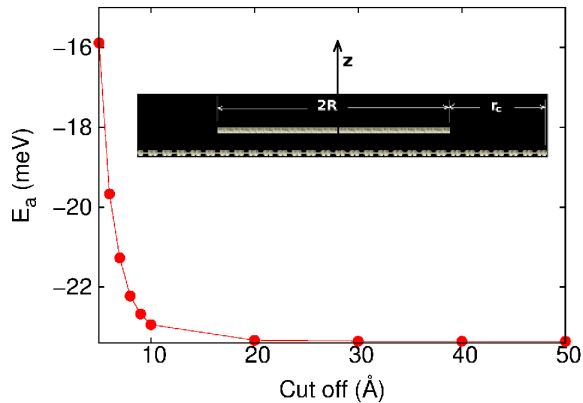}}
  \end{center}
  \caption{(Color online) The interlayer potential energy calculated from the registry-dependent potential with different cut-off. The $y$-axis ($E_{a}$) is the total interlayer energy per atom. In this calculation, the radius of the top graphene layer is $R=100$~{\AA}. The twisting angle $\theta=0.0$, i.e AB-stacking BLG. For a cut-off $r_{c}=25$~{\AA}, the variation in the energy is on the order of $10^{-3}$~{meV}. Inset displays the cross section of a twisting BLG, where $R$ is the radius and $r_{c}$ is the boundary region.}
  \label{fig_rcut_registry}
\end{figure}
In this paper, we calculate the interlayer energy in the twisting BLG. The interlayer interaction is described by either the registry-dependent potential or the Lennard-Jones potential. We find that the interlayer energy does not depend on the twisting angle, except in the boundary regions $\theta\approx 0$, $60^{\circ}$, where the interlayer energy is proportional to the square of the twisting-related arc length $S=R\theta$. We explain these results by averaging the energy distribution in a single atom (SA) model. The twisting angle-dependence of the interlayer energy actually provides an explanation for the diverse twisting angles observed in the experiment.

The present paper is organized as follows. In Sec.~II, after a brief description of the structure, we present the detailed results based on the registry-dependent potential. Sec.~III is devoted to the simulation results for the Lennard-Jones potential. The paper ends with a brief summary in Sec.~IV.

\section{results and discussion for registry-dependent potential}
\begin{figure}[htpb]
  \begin{center}
    \scalebox{1.0}[1.0]{\includegraphics[width=8cm]{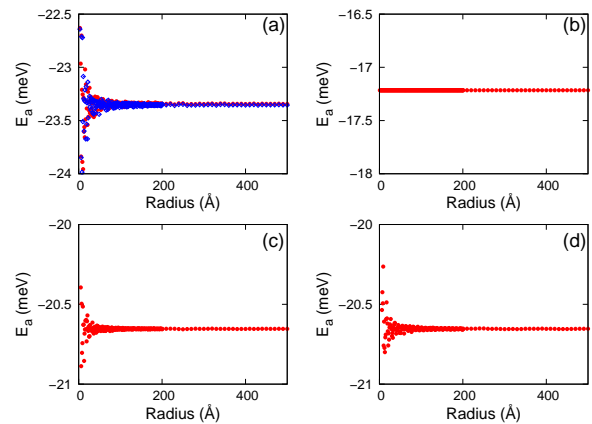}}
  \end{center}
  \caption{(Color online) The saturation of the energy per atom with increasing radius for different twisting angles $\theta$. (a) $\theta=0$, i.e AB-stacking BLG. Solid and open dots correspond to cut-off $r_{c}=$ 25 and 50~{\AA}. (b) $\theta=60^{\circ}$, i.e AA-stacking BLG. The energy per atom in the AA-stacking BLG is radius independent. (c) $\theta=21.8^{\circ}$, i.e the first commensurable angle. (d) $\theta=13.2^{\circ}$, i.e the second commensurable angle.}
  \label{fig_radius_registry}
\end{figure}
We start with the AB-stacking BLG. The $xy$ axes sit in the bottom layer. In the AB-stacking, there is a carbon atom from the top layer sitting on the head of a carbon atom from the bottom layer. The $z$ axis comes across this pair of atoms, and the origin of the coordinate system is set on the carbon atom in the bottom layer. The bottom layer is infinite large. This `infinite' is numerically realized by ensuring that the radius of the bottom layer is always larger than $\left (R+r_{c}\right )$, where $R$ is the radius of the top graphene layer and $r_{c}$ is the cut-off in the interlayer potential (see inset in Fig.~\ref{fig_rcut_registry}). The top layer is then twisted about the $z$ axis for an arbitrary twisting angle $\theta$. In particular, the AB-stacking BLG has a twisting angle $\theta=0$, and the AA-stacking BLG has $\theta=60^{\circ}$.

Previous research has shown the limitation of the Lennard-Jones potential in the description of the interlayer interaction in BLG.\cite{GirifalcoLA} Particularly, it only gives less than $10\%$ of the twisting or shearing energy in BLG. To improve the situation, the Lennard-Jones potential is extended to be registry-dependent.\cite{KolmogorovAN2000,KolmogorovAN2005,LebedevaIVpccp} In our calculation, we have employed the latest version of the registry-dependent potential developed by Lebedeva {\it et.al} in 2011, which has been succeeded in predicting the energy barrier for a relative translation of the two graphene layers in BLG.\cite{LebedevaIVpccp,PopovAM} In this model, the interaction between two atoms from adjacent graphene layers is described by following formula:
\begin{eqnarray}
V\left(r\right) & = & A\left(\frac{z_{0}}{r}\right)^{6}+Be^{-\alpha\left(r-z_{0}\right)}\nonumber\\
&+&C\left(1+D_{1}\rho^{2}+D_{2}\rho^{4}\right)e^{-\lambda_{1}\rho^{2}}e^{-\lambda_{2}\left(z^{2}-z_{0}^{2}\right)},
\label{eq_rdp}
\end{eqnarray}
where $r$ is the distance between two atoms, and $\rho^{2} = r^{2}-z^{2}$. The parameters are as follows: $A=-10.510$~{meV}, $z_{0}=3.34$~{\AA}, $B=11.652$~{meV}, $\alpha=4.16$~{\AA$^{-1}$}, $C=35.883$~{meV}, $D_{1}=-0.86232$~{\AA$^{-2}$}, $D_{2}=0.10049$~{\AA$^{-4}$}, $\lambda_{1}=0.48703$~{\AA$^{-2}$}, $\lambda_{2}=0.46445$~{\AA$^{-2}$}. The interaction cut off is $r_{\rm cut}=25$~{\AA}. These parameters were fitted to both ab-initio (DFT-D) calculated results and some experimental results. The parameters for the isotropic part (A, B, and $\alpha$) were fitted to the experimental value of binding energy, interlayer space and the c-axis compressibility for graphite. The parameters in the anisotropic part determine the dependence of the interlayer potential on the in-plane relative displacement of graphene layers.
\begin{figure}[htpb]
  \begin{center}
    \scalebox{1.0}[1.0]{\includegraphics[width=8cm]{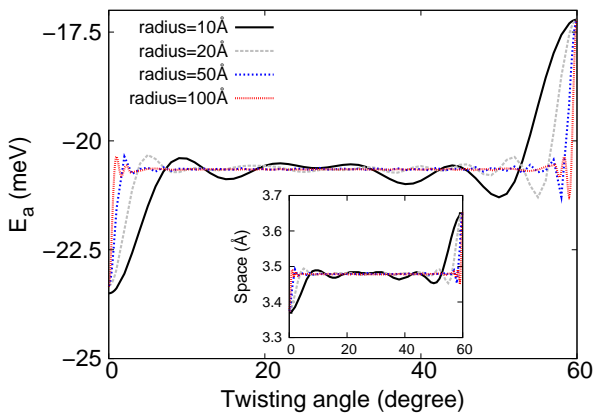}}
  \end{center}
  \caption{(Color online) The energy per atom versus twisting angle $\theta$. The whole curve is symmetric about $\theta=60^{\circ}$ due to the mirror symmetry between the two nonequivalent atoms in graphene, so $\theta\in[60,120]^{\circ}$ is not shown. Inset shows the interlayer space versing twisting angle $\theta$.}
  \label{fig_theta_registry}
\end{figure}
 These anisotropic parameters were obtained by fitting to the ab-initio calculated results for the relative energies for the AA and intrinsic unstable stackings, and the frequency of the interlayer shearing mode. The interlayer binding energy mainly depends on the isotropic parameters (A, B, and $\alpha$); while the interlayer shearing energy is mostly controlled by the other parameters in the anisotropic potential.

For different twisting angle, the structure is relaxed to the energy minimum state with boundary atoms fixed. In particular, the interlayer space is also relaxed. We then calculate the interlayer energy per atom in this optimized structure, $E_{a}=E_{\rm tot}/N_{\rm top}/2$, where $E_{\rm tot}$ is the total interlayer energy between the two graphene layers and $N_{\rm top}$ is the total carbon atoms in the top layer. A factor of 2 is to divide the pair energy into each atom. We note that $E_{a}$ here is different from the binding energy defined in Refs.~\onlinecite{LebedevaIVpccp,PopovAM} by a factor of 2. We first examine effects from the cut off in the potential. Fig.~\ref{fig_rcut_registry} shows the energy per atom calculated with different cut off in AB-stacking BLG. In this calculation, the radius of the top layer is 100~{\AA}. The variation is on the order of $10^{-3}$~meV for cut-off 25~{\AA}. In the following calculation for the registry-dependent potential, we use a cut-off 25~{\AA}, so that our calculation precision is $10^{-3}$~meV.

Fig.~\ref{fig_radius_registry} shows the convergence of $E_{a}$ with increasing radius ($R$) in BLG with various twisting angles. The twisting angles are 0.0, $60^{\circ}$, $21.8^{\circ}$, and $13.2^{\circ}$ in panels from (a) to (d). These angles correspond to the commensurable angles from Santos's equation\cite{Santos} $\theta=\arccos ((3i^2+3i+0.5)/(3i^2+3i+1))$ with integers $i=$0, 1, 2, and $+\infty$. For all panels, the energy saturates at radius around 100~{\AA}, i.e the oscillation in the energy for a system with $r>100$~{\AA} is smaller than our calculation precision ($10^{-3}$~{meV}). In panel (a), i.e AB-stacking BLG, $E_{a}$ converges to a saturate value of -23.345~meV.
\begin{figure}[htpb]
  \begin{center}
    \scalebox{1.0}[1.0]{\includegraphics[width=8cm]{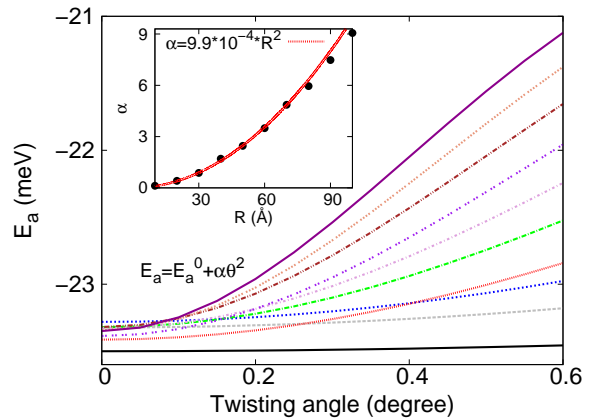}}
  \end{center}
  \caption{(Color online) The energy per atom versus twisting angle around $\theta=0$. For curves from bottom to top, the radius increases from 10 to 100~{\AA}. All curves can be well fitted to $E_{a}=E_{a}^{0}+\alpha \theta^{2}$, where $E_{a}^{0}$ is the value at $\theta=0$. $\alpha$ is the fitting parameter. Inset shows the radius-dependence of $\alpha$. $\alpha$ is fitted to $\alpha=9.9\times 10^{-4}R^{2}$.}
  \label{fig_theta_0_registry}
\end{figure}
 In this calculation, two different cut-off have been used. The data corresponding to $r_{c}=25$~{\AA} are represented by solid dots. Results for $r_{c}=50$~{\AA} are displayed by open symbols. The saturate value is the same in both case. It further confirms that $r_{c}=25$~{\AA} is suitable. In panel (b), i.e AA-stacking BLG, $E_{a}$=-17.215~meV is radius independent. The energy difference between these two stacking style is 6.130~meV. In panels (c) and (d), $E_{a}$ converges to the same saturate values of -20.654~meV. It is interesting that the energy in a BLG with different commensurable angles are the same. It indicates that the commensurable twisting angle may not correspond to the energy minimum state. Hence, from the energy point of view, the twisting angle existing in the experiment is not necessarily the commensurable angle.

Fig.~\ref{fig_theta_registry} further confirms that $E_{a}$ does not depend on the twisting angle in a wide angle range $\theta\in[10, 50]^{\circ}$, except in the two boundary regions ($\theta\approx 0$, $60^{\circ}$). For small radius, there is some obvious oscillation in the range of $\theta\in[10,50]^{\circ}$. With increasing radius, this oscillation amplitude decays and becomes indistinguishable after $R>50$~{\AA}. It indicates that the oscillation is actually due to the size effect, and is not related to the twisting angle. Similar size effect also exists in a recent work by Shibuta and Elliott.\cite{ShibutaY} The inset in Fig.~\ref{fig_theta_registry} shows the optimized interlayer space of the BLG with different twisting angle, which looks quite similar as the energy curve. Our calculation predicts that the interlayer space is not sensitive to the twisting angle in a large angle range.

\begin{figure}[htpb]
  \begin{center}
    \scalebox{1.0}[1.0]{\includegraphics[width=8cm]{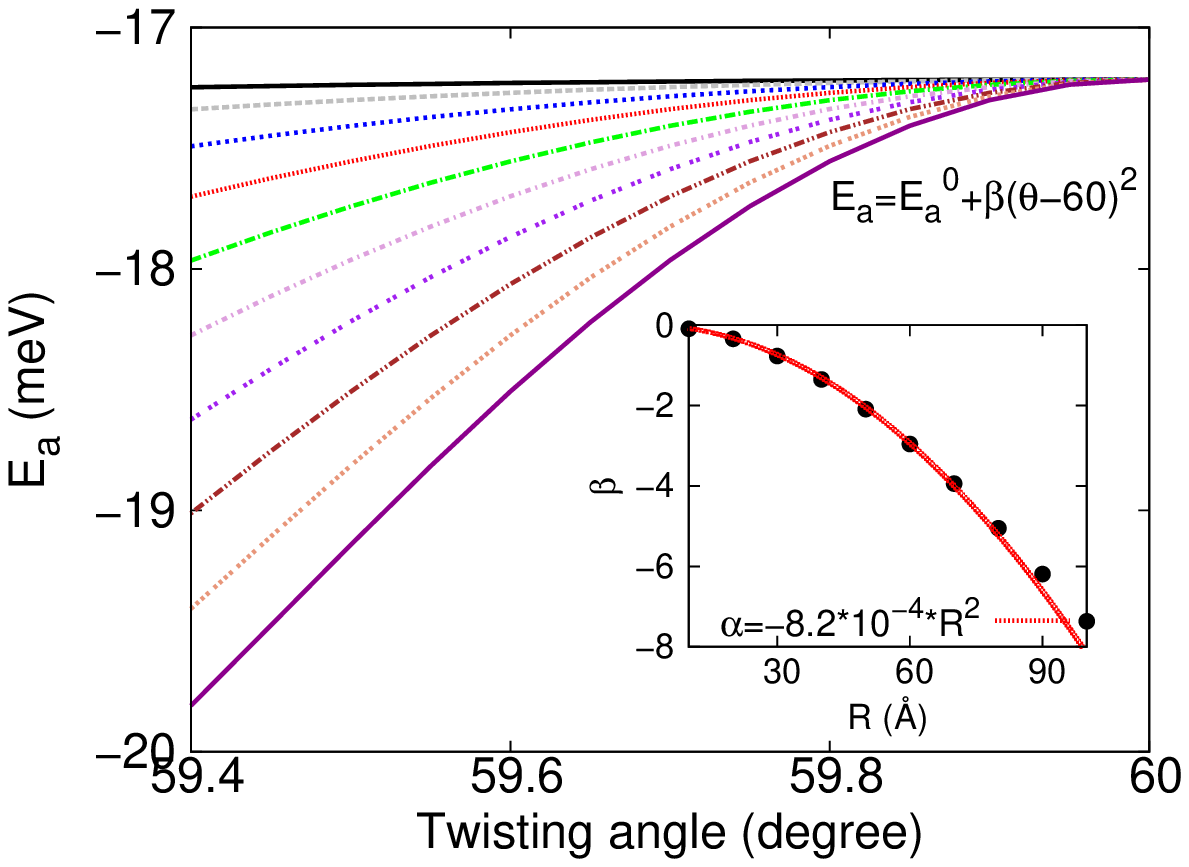}}
  \end{center}
  \caption{(Color online) The energy per atom versus twisting angle around $\theta=60^{\circ}$. For curves from top to bottom, the radius increases from 10 to 100~{\AA}. All curves can be well fitted to $E_{a}=E_{a}^{0}+\beta (\theta-60)^{2}$, where $E_{a}^{0}$ is the value at $\theta=60^{\circ}$. $\beta$ is the fitting parameter. Inset shows the radius-dependence of $\beta$. $\beta$ is fitted to $\beta=-8.2\times 10^{-4}R^{2}$.}
  \label{fig_theta_60_registry}
\end{figure}

In Fig.~\ref{fig_theta_registry}, although the curve is a platform in a wide angle range, we can see that it changes sharply around the two boundaries $\theta\approx0$ and $60^{\circ}$. These two regions are zoomed in in Fig.~\ref{fig_theta_0_registry} and Fig.~\ref{fig_theta_60_registry}. Fig.~\ref{fig_theta_0_registry} shows that $E_{a}$ can be fitted to $E_{a}=E_{a}^{0}+\alpha \theta^{2}$, where $E_{a}^{0}$ is the value at $\theta=0$. $\alpha$ is the fitting parameter and is radius-dependent. Inset shows the radius-dependence of $\alpha$, which is fitted to $\alpha=9.9\times 10^{-4}R^{2}$.
 Hence, around $\theta=0$, we have $E_{a}=E_{a}^{0}+9.9\times 10^{-4}R^{2}\theta^{2}\equiv E_{a}^{0}+9.9\times 10^{-4}S^{2}$, where $S=R\theta$ is the twisting-related arc length. Similar results are observed for $\theta\approx60^{\circ}$ as shown in Fig.~\ref{fig_theta_60_registry}, where we obtain $E_{a}=E_{a}^{60}-8.2\times 10^{-4}R^{2}(\theta-60)^{2}$. $E_{a}^{60}$ is the value at $\theta=60^{\circ}$.

\begin{figure}[htpb]
  \begin{center}
    \scalebox{1.0}[1.0]{\includegraphics[width=8cm]{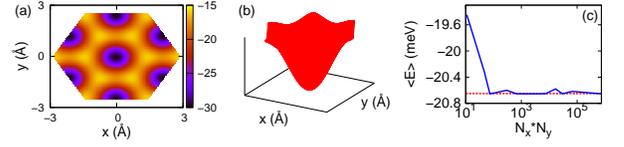}}
  \end{center}
  \caption{(Color online) The registry-dependent potential between a single atom and an infinite large graphene sheet. The single atom is on top of the graphene at a fixed distance $z=3.478$~{\AA}. $x$ and $y$ axes in the figure are the other two coordinates of the single atom. (a) shows six-fold symmetry in the energy, due to the hexagon structure of graphene. (b) shows only one valley of the energy. (c) shows the convergence of the average of the energy over $xy$ area with increasing grid points. The dashed line (red online) depicts the platform value of the energy per atom in Fig.~\ref{fig_theta_registry}.}
  \label{fig_lj_one_atom_registry}
\end{figure}

In the above, we have presented the angle dependence for the interlayer potential of the twisting BLG. The remainder of this section will be devoted to explaining the origin for this angle dependence. Let's consider a single carbon atom on top of an infinite graphene sheet. The distance from this single atom to the graphene is fixed to be 3.478~{\AA}, which is the platform value in the inset of Fig.~\ref{fig_theta_registry}. We refer to such system as single atom (SA) model. The interlayer energy per atom in the SA model is calculated by $E_{\rm SA}=E_{\rm tot}/2$, where $E_{\rm tot}$ is the total interlayer energy between the single atom and the infinite graphene. Fig.~\ref{fig_lj_one_atom_registry}~(a) shows the energy distribution of the SA model. The $x$ and $y$ axes are the $xy$ position of the single atom. There are translational and six-fold rotational symmetries in the energy distribution corresponding to the honeycomb structure of the graphene sheet. Fig.~\ref{fig_lj_one_atom_registry}~(b) shows a three-dimensional plot of the energy distribution within one hexagonal area. The hexagon is formed by six carbon atoms. When the single atom is on top of the hexagon corner (position A), $E_{\rm SA}$ reaches the maximum value of $E_{\rm SA}=E_{A}=-16.045$~meV. When the single atom is on top of the hexagon center (position B), $E_{\rm SA}$ achieves the minimum value of $E_{\rm SA}=E_{B}=-29.810$~meV. We note that the origin of the coordinate system here is set to the center of the hexagon in the bottom layer.

For an arbitrary twisting angle, the positions of all atoms are arbitrarily distributed between position A and B. The energy per atom of such a system should be equivalent to the average of the energy in Fig.~\ref{fig_lj_one_atom_registry}~(b) over the $xy$ area within one hexagon:
\begin{eqnarray}
<E>=\frac{\int E_{\rm SA}(x,y) dxdy}{\int dxdy}.
\label{eq_average}
\end{eqnarray}
We use a rectangular grid to partition the $xy$ area, with $N_{x}$ and $N_{y}$ points in the $x$ and $y$ directions. The number of total grid points is $N_{x}\times N_{y}$. Fig.~\ref{fig_lj_one_atom_registry}~(c) shows the convergence of the average with increasing grid points $N_{x}\times N_{y}$. The average converges to a saturate value, which is exactly the energy per atom in Fig.~\ref{fig_theta_registry} (the dashed red line). \textit{We propose a mapping between twisting BLG and the SA model: different twisting angle in the BLG corresponds to different grid type in the integration of Eq.~(\ref{eq_average}).} According to the Riemann-Lebesgue theorem,\cite{BochnerS} the integration in Eq.~(\ref{eq_average}) exists and does not depend on the grid type, because the integral function is bounded and smooth everywhere. From the mapping, this theorem tells us that the energy per atom is the same for twisting BLG with twisting angle $\theta$ in a large angle range. It explains why $E_{a}$ does not depend on the twisting angle in a wide range.

It should be note that the SA model can not be applied to explain the two boundary regions $\theta\approx0$ and 60$^{\circ}$. It is because the SA model depends on the interlayer space, while the space is sensitive to the twisting angle in these two boundary regions around $\theta=0$ and 60$^{\circ}$. As will be shown bellow, if the interlayer space is independent of the twisting angle, then the SA model will succeed for all twisting angles, including $\theta\approx0$ and 60$^{\circ}$.

{\it Summary for this section.} Using the registry-dependent potential, we have shown that the energy of the twisting BLG is insensitive to the twisting angle in a large angle range, which can be analyzed by the SA model. It illustrates that, from the energy point of view, there is no favorable twisting angles for the twisting BLG with $\theta\in[10, 50^{\circ}]$. Particularly, Moire pattern, i.e structure with a commensurate twisting angle does not correspond to the energy minimum structure.

\begin{figure}[htpb]
  \begin{center}
    \scalebox{1.0}[1.0]{\includegraphics[width=8cm]{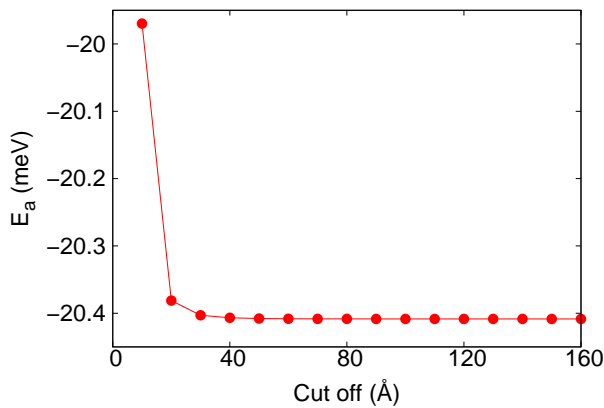}}
  \end{center}
  \caption{(Color online) The interlayer potential energy calculated from Lennard-Jones potential with different cut-off. The $y$-axis ($E_{a}$) is the energy per atom. In this calculation, the radius of the top layer is 100~{\AA}. The twisting angle $\theta=21.8^{\circ}$. For a cut-off distance of $r_{c}=100$~{\AA}, the variation in the energy is on the order of $10^{-5}$~{meV}.}
  \label{fig_rcut}
\end{figure}
\begin{figure}[htpb]
  \begin{center}
    \scalebox{1.0}[1.0]{\includegraphics[width=8cm]{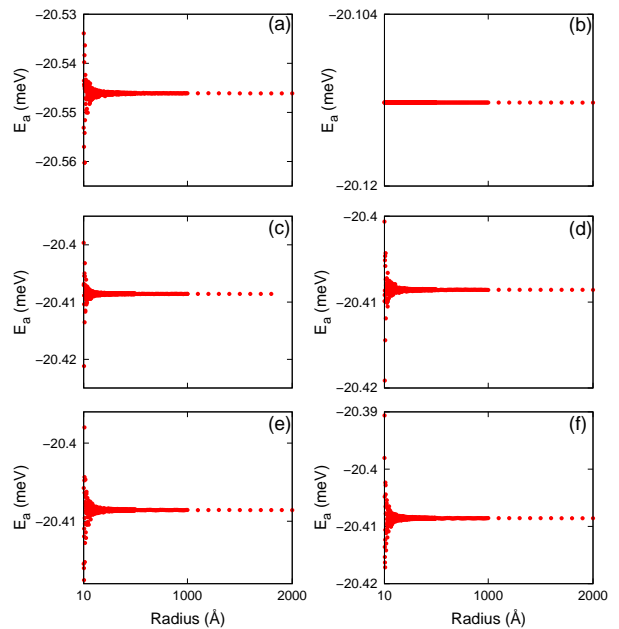}}
  \end{center}
  \caption{(Color online) The saturation of the energy per atom with increasing radius for different twisting angles $\theta$. (a) $\theta=0$, i.e AB-stacking BLG. (b) $\theta=60^{\circ}$, i.e AA-stacking BLG. The energy per atom in the AA-stacking BLG is radius independent. (c) $\theta=21.8^{\circ}$, i.e the first commensurable angle. (d) $\theta=23^{\circ}$. (e) $\theta=13.2^{\circ}$, i.e the second commensurable angle. (f) $\theta=11^{\circ}$.}
  \label{fig_radius}
\end{figure}
\section{results and discussion for Lennard-Jones potential}
Although Lennard-Jones potential is insufficient for the description of interlayer interaction in BLG, in this section, for the convenience of comparison, we present simulation results based on the Lennard-Jones potential, $V(r)=4\epsilon ((\sigma/r)^{12}-(\sigma/r)^{6})$, with $\epsilon=2.5$ meV and $\sigma=3.37$~{\AA}. The two parameters $\sigma$ and $\epsilon$ are fitted to experimental values for the interlayer space and the phonon dispersion along $\Gamma A$ direction in three-dimensional graphite.\cite{JiangJW} Usually, the cut-off in the Lennard-Jones potential is set to be around 10~{\AA}. Fig.~\ref{fig_rcut} shows the energy per atom for a BLG with top layer radius as 100~{\AA} and the twisting angle $\theta=21.3^{\circ}$. It shows that the variation is still on the order of 0.1~meV for cut-off 10~{\AA}. This variation is too large for present study. For a cut-off 100~{\AA}, the variation in the $E_{a}$ is on the order of $10^{-5}$~meV. In the following calculation, we use a cut-off 100~{\AA} for Lennard-Jones potential, so that our calculation precision is $10^{-5}$~meV. The required precision is much higher for Lennard-Jones potential than the registry-dependent potential, because the Lennard-Jones potential gives much weaker twisting energy. Without losing universality, we use the same interlayer space, 3.35~{\AA}, for BLG with all twisting angles in this section.

\begin{figure}[htpb]
  \begin{center}
    \scalebox{1.0}[1.0]{\includegraphics[width=8cm]{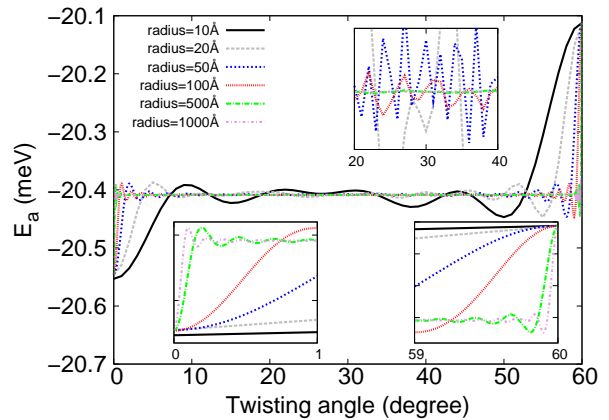}}
  \end{center}
  \caption{(Color online) The energy per atom versus twisting angle $\theta$. The top inset shows the close-up of the middle region $\theta\in[20,40]^{\circ}$, where curves for large radius of 500 and 1000~{\AA} are indistinguishable. Two bottom insets show the close-up of the boundary regions $\theta\approx 0$ amd $60^{\circ}$.}
  \label{fig_theta}
\end{figure}
Fig.~\ref{fig_radius} shows the convergence of $E_{a}$ with increasing radius ($R$) in BLG at various twisting angles. The twisting angles are 0.0, $60^{\circ}$, $21.8^{\circ}$, $23.0^{\circ}$, $13.2^{\circ}$, and $11.0^{\circ}$ in panels from (a) to (f). Panels (d) and (f) correspond to two arbitrary twisting angles, while all other four panels correspond to commensurable angles. For all panels, the energy saturates at radius around 500~{\AA}. In panel (a), AB-stacking BLG, $E_{a}$ converges to a saturate value of -20.54613~meV. In panel (b), AA-stacking BLG, the saturate value is -20.11223~meV. The energy difference between these two stacking styles is 0.43390~meV, which is one order smaller than the value from the registry-dependent potential in the previous section.
\begin{figure}[htpb]
  \begin{center}
    \scalebox{1.0}[1.0]{\includegraphics[width=8cm]{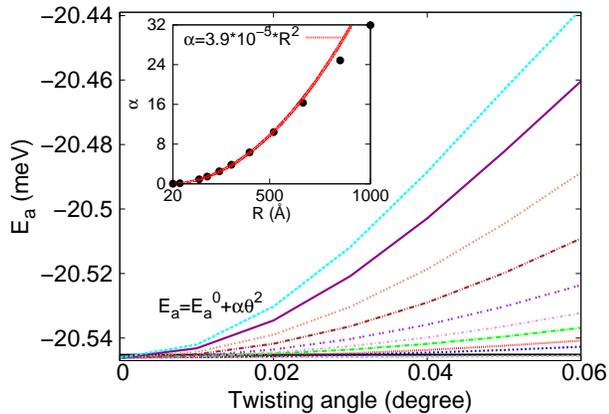}}
  \end{center}
  \caption{(Color online) The energy per atom versus twisting angle around $\theta=0$. For curves from bottom to top, the radius increases as 20, 55, 150, 190, 250, 310, 400, 520, 665, 850, and 1000~{\AA}. All curves can be well fitted to $E_{a}=E_{a}^{0}+\alpha \theta^{2}$, where $E_{a}^{0}$ is the value at $\theta=0$. $\alpha$ is the fitting parameter. Inset shows the radius-dependence of $\alpha$. $\alpha$ is fitted to $\alpha=3.9\times 10^{-5}R^{2}$.}
  \label{fig_theta_0}
\end{figure}
 In all other four panels, $E_{a}$ converges to the same saturate values of -20.40857~meV. It becomes interesting that the energy in a BLG with commensurable angles shown in panels (c) and (e) are the same as that in BLG with arbitrary twisting angles shown in panels (d) and (f). It illustrates that the commensurable twisting angle does not correspond to the energy minimum state. Hence, from the energy point of view, the twisting angle observed in the experiment is not necessarily the commensurable angle. Instead, it can be an arbitrary angle. We have observed similar phenomenon in previous section, where the interlayer interaction is described by the registry-dependent potential; so this finding does not depend on the potential type.

Fig.~\ref{fig_theta} shows that $E_{a}$ does not depend on the twisting angle in a wide angle range, except in the two boundary regions ($\theta\approx 0$, $60^{\circ}$). For small radius, there is some obvious oscillation in the range of $\theta\in[10,50]^{\circ}$. With increasing radius, this oscillation amplitude decreases and becomes indistinguishable after $R>500$~{\AA}. It indicates that the oscillation is actually due to the size effect, and is not related to the twisting angle. The top inset shows the close-up of the curve in the angle range of $\theta\in[20,40]^{\circ}$, where the two results for large radius of 500 and 1000~{\AA} are indistinguishable. According to the energy minimum condition criteria, these results imply that the twisting angle can be an arbitrary value in this range. Two bottom insets show the close-up of the boundary regions $\theta\approx 0$ amd $60^{\circ}$.

\begin{figure}[htpb]
  \begin{center}
    \scalebox{1.0}[1.0]{\includegraphics[width=8cm]{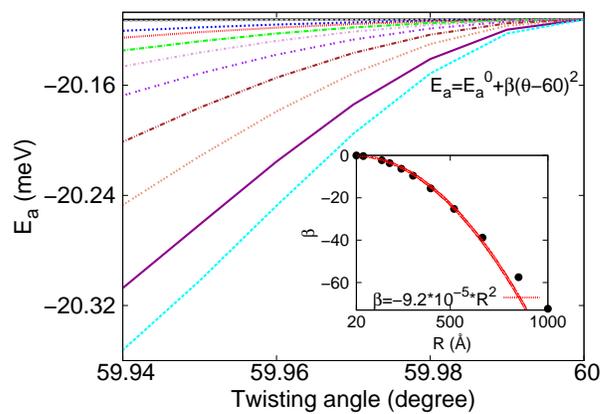}}
  \end{center}
  \caption{(Color online) The energy per atom versus twisting angle around $\theta=60^{\circ}$. For curves from top to bottom, the radius increases as 20, 55, 150, 190, 250, 310, 400, 520, 665, 850, and 1000~{\AA}. All curves can be well fitted to $E_{a}=E_{a}^{0}+\beta (\theta-60)^{2}$, where $E_{a}^{0}$ is the value at $\theta=60^{\circ}$. $\beta$ is the fitting parameter. Inset shows the radius-dependence of $\beta$. $\beta$ is fitted to $\beta=-9.2\times 10^{-5}R^{2}$.}
  \label{fig_theta_60}
\end{figure}
These two boundary regions are further zoomed in in Fig.~\ref{fig_theta_0} and Fig.~\ref{fig_theta_60}. Fig.~\ref{fig_theta_0} shows that $E_{a}$ can be fitted to $E_{a}=E_{a}^{0}+\alpha \theta^{2}$, where $E_{a}^{0}$ is the value at $\theta=0$. $\alpha$ is the fitting parameter and is radius-dependent. Inset shows the radius-dependence of $\alpha$, which is fitted to $\alpha=3.9\times 10^{-5}R^{2}$. Hence, around $\theta=0$, we have $E_{a}=E_{a}^{0}+3.9\times 10^{-5}R^{2}\theta^{2}\equiv E_{a}^{0}+3.9\times 10^{-5}S^{2}$, where $S=R\theta$ is the twisting-related arc length. Similar results are observed for $\theta\approx60^{\circ}$ as shown in Fig.~\ref{fig_theta_60}, where we obtain $E_{a}=E_{a}^{60}-9.2\times 10^{-5}R^{2}(\theta-60)^{2}$. $E_{a}^{60}$ is the value at $\theta=60^{\circ}$.

\begin{figure}[htpb]
  \begin{center}
    \scalebox{1.0}[1.0]{\includegraphics[width=8cm]{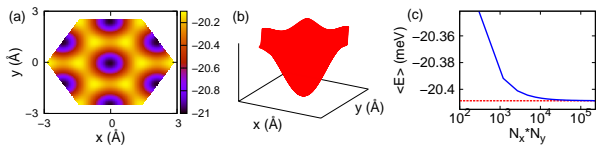}}
  \end{center}
  \caption{(Color online) The Lennard-Jones potential between a single atom and an infinite large graphene sheet. The single atom is on top of the graphene at a fixed distance $z=c=3.35$~{\AA}. (a) shows six-fold symmetry in the energy, due to the hexagon structure of graphene. (b) shows only one valley of the energy. (c) shows the convergence of the average of the energy over $xy$ area with increasing grid points. The dashed line (red online) depicts the platform value of the energy per atom in Fig.~\ref{fig_theta}.}
  \label{fig_lj_one_atom}
\end{figure}

We are now applying the SA model to analyze these simulation results. In the SA model, the distance from the single atom to the graphene is fixed to be 3.35~{\AA}. We calculate the energy per atom in this SA model. Fig.~\ref{fig_lj_one_atom}~(a) shows the energy distribution of the SA model. Fig.~\ref{fig_lj_one_atom}~(b) shows a three-dimensional plot of the energy distribution within one hexagonal area. When the single atom is on top of the hexagon corner (position A), $E_{\rm SA}$ reaches the maximum value of $E_{\rm SA}=E_{A}=-20.11223$~meV. When the single atom is on top of the hexagon center (position B), $E_{\rm SA}$ achieves the minimum value of $E_{\rm SA}=E_{B}=-20.98000$~meV.

In the AB-stacking BLG, half of the atoms are located at position A and the other half atoms are located at position B. As a result, we get $E_{a}=\left( E_{A}+E_{B}\right)/2=-20.54612$~meV, which agrees quite well with the saturate value from Fig.~\ref{fig_radius}~(a). In the AA-stacking BLG, all atoms are at position A, so $E_{a}=E_{A}=-20.11223$~{meV}, which is exactly the same as the value from Fig.~\ref{fig_radius}~(b). For an arbitrary twisting angle, the positions of all atoms are arbitrarily distributed between position A and B. The energy per atom of such a system should be equivalent to the average of the energy in Fig.~\ref{fig_lj_one_atom}~(b) over the $xy$ area within one hexagon following Eq.~(\ref{eq_average}). Fig.~\ref{fig_lj_one_atom}~(c) shows the convergence of the average with increasing grid points $N_{x}\times N_{y}$. The average converges to a saturate value, which is exactly the energy per atom in Fig.~\ref{fig_theta} (the dashed red line).

We note that the interlayer space here has been kept the same for all twisting angles, so the SA model (with space 3.35~{\AA}) can be used for all twisting angles, including the two boundary regions $\theta\approx0$ and $60^{\circ}$. The energy curve in Fig.~\ref{fig_lj_one_atom}~(b) within the vicinity of the valley or the peak can be well described by functions $E_{a}=E_{\rm min}+1.16r^{2}-0.44r^{4}$ and $E_{a}=E_{\rm max}-0.66r^{2}+0.53r^{3}$, respectively, where the distance $r$ is with respect to the minimum point at the valley or the maximum point in the peak.

For a twisting BLG, if the twisting angle $\theta=0$, then half of the carbon atoms in the top layer are located at position B, i.e the minimum point of the valley; while the other half of atoms are at position A, i.e the maximum point of the peak. If the twisting angle is small but close to 0, i.e $\theta\approx 0$, half of the carbon atoms in the top layer deviate slightly from the valley while the other half of atoms deviate slightly from the maximum point of the peak. The deviation of each atom is $s\approx r*\theta$, where $r$ is the radial position of the atom. The edge atoms have the largest deviation $S\approx R\theta$, where $R$ is the radius of the top graphene layer.

It is clear that the energy per atom for the first half atoms can also be calculated by averaging of Eq.~(\ref{eq_average}) over a circular area around the valley with radius $S$ in the $xy$ plane, yielding $<E>=E_{\rm min}+1.76\times10^{-4} (R\theta)^{2}-1.4\times 10^{-8}(R\theta)^{4}$, where $\theta$ is in the unit of degree. For the other half atoms, the energy per atom can be obtained by doing average of Eq.~(\ref{eq_average}) over a circular area around the peak with radius $S$ in the $xy$ plane. The integral result is $<E>=E_{a}^{0}-1.01\times10^{-4} (R\theta)^{2}+1.13\times10^{-6} (R\theta)^{3}$. As a result, the energy per atom for the twisting BLG with small twisting angle is $<E>\approx (E_{\rm min}+E_{\rm max})/2+3.8\times10^{-5} (R\theta)^{2}$, where the coefficient $3.8\times10^{-5}$ agrees well with the results ($3.9\times10^{-5}$) in Fig.~\ref{fig_theta_0}. Similarly, for twisting BLG with twisting angle $\theta$ around $60^{0}$, the energy of all atoms deviate slightly from the maximum point of the peak, so the energy per atom from the integral of Eq.~(\ref{eq_average}) is $<E>\approx E_{\rm max}-1.01\times10^{-4} (R\theta)^{2}$, with the coefficient $-1.01\times10^{-4}$ quite close to the results ($9.2\times10^{-5}$) in Fig.~\ref{fig_theta_60}. Now it also becomes clear that the discrepancy in large radius limit between the fitting curve and the calculated data in Fig.~\ref{fig_theta_0} and Fig.~\ref{fig_theta_60} is due to the nonlinear potential energy in higher order terms of $(R\theta)$.

\section{conclusion}
In conclusion, using both registry-dependent potential and Lennard-Jones potential, we have shown that the interlayer energy in a twisting BLG is independent of the twisting angle except in the boundary regions $\theta\approx 0$, $60^{\circ}$. In these two boundaries, the interlayer energy is related to the square of the twisting-related arc length (S), $E_{a}=E_{a}^{0}+cS^{2}$, with a constant coefficient $c$. These observations are well explained by averaging the energy distribution in the SA model, where a BLG with a particular twisting angle is mapped to a special grid type in the partition of the integration area in the averaging. The energy distribution function in the SA model is so smooth that its integration does not depend on the grid type, leading to the twisting angle independence of the interlayer potential in a twisting BLG. Our theoretical results indicate that there is no preference for a commensurable twisting angle in theBLG sample, according to the minimum energy condition criterion. These findings provide a possible explanation for the diverse twisting angles observed in the experiment.

\textbf{Acknowledgements} We would like to thank the innominate referee for suggesting the registry-dependent interlayer potential. The work is supported in part by the Grant Research Foundation (DFG).

\end{document}